# Generalization of Zernike polynomials for regular portions of circles and ellipses

**Rafael Navarro[1,*] José L. López[2], José A. Díaz[3], and Ester Pérez Sinusía[4]**

1. ICMA, Consejo Superior de Investigaciones Científicas & Universidad de Zaragoza. Facultad de Ciencias, P. Cerbuna 12, 50009 Zaragoza, Spain
2. Departamento de Ingeniería Matemática e Informática, Universidad Pública de Navarra. 31016 Pamplona. Navarra. Spain
3. Departamento de Óptica, Universidad de Granada, 18071 Granada, Spain
4. Departamento de Matemática Aplicada, Universidad de Zaragoza. Zaragoza, Spain

**Corresponding author:** rafaelnb@unizar.es

## ABSTRACT

Zernike polynomials are commonly used to represent the wavefront phase on circular optical apertures, since they form a complete and orthonormal basis on the unit circle. Here, we present a generalization of this Zernike basis for a variety of important optical apertures. On the contrary to ad hoc solutions, most of them based on the Gram-Smith orthonormalization method, here we apply the diffeomorphism (mapping that has a differentiable inverse mapping) that transforms the unit circle into an angular sector of an elliptical annulus. In this way, other apertures, such as ellipses, rings, angular sectors, etc. are also included as particular cases. This generalization, based on in-plane warping of the basis functions, is unique, and what is more important, it guarantees a reasonable level of invariance of the mathematical properties and the physical meaning of the initial basis functions. Both, the general form and the explicit expressions for most common, elliptical and annular apertures are provided.

## I. INTRODUCTION

The problem of finding complete orthonormal systems to represent functions defined on finite supports with a given geometry appears in many areas of Physics and Engineering. In particular, Zernike circle polynomials [ 1 ] are widely used to represent optical path differences (phase differences or wave aberrations) in wavefronts, or even the sag of optical surfaces (such as the human cornea [2, 3]) as they are well adapted to the circular shape of a majority of conventional optical systems. There is an infinite number of possible systems, but Zernike polynomials (ZPs) (or lineal combinations of them [4]) show important advantages and interesting properties. Among these properties, ZPs permit to establish a link with the traditional Seidel theory of aberrations [5], which is based on a third order Taylor series expansion, and with further extensions of the Seidel theory to 5th order, etc. On the one hand, the monomials of the Taylor series have a clear physical meaning as they represent different types of aberrations (defocus, astigmatism, coma, etc.) but are not orthogonal, which limits both theoretical and practical developments. On the other hand, higher order Zernike polynomials contain lower order terms, as a necessary balance to get zero average [6]. As a result of this cross-talk between higher and lower orders, the link of the ZPs to the Seidel theory is not evident. Nevertheless, the theoretical and practical advantages of orthonormal polynomials, make that ZPs became the standard way to describe the phase of wavefronts [7] (or the wave aberration or optical path differences) in many fields ranging from atmospheric optics [8], optical design and testing [9] or visual optics (the ANSI Z80.28 standard for reporting aberrations in the human eye is based on ZPs). Even though the circle is the most common optical aperture, there are other geometries, such as the annular pupils in large telescopes [10] or rectangular in anamorphic systems, etc. Furthermore, even in the case of circular apertures, the effective pupil becomes elliptical for off-axis field angles [11]. The eccentricity of the ellipse increases with field angle, and can reach high values for wide angle lenses. The importance of this problem motivated the development of a series of ad hoc solutions, most of them based on the Gram-Smith (G-S) method to obtain orthonormal basis on different types of apertures [12, 13] such as ellipses [11], rectangles [14], annuli [15], circular sectors [16], etc. In some particular cases, such as rectangles, Legendre [14] or Chebyshev polynomials [17] were also proposed, but their integration within Seidel or Zernike theoretical frameworks is difficult. An extensive catalogue of polynomial basis functions on different types of apertures can be found in [18]. The G-S method has several advantages but also important drawbacks. The main advantage is that it is a quite general linear method, so that the resulting basis functions are linear combinations of the initial Zernike circle polynomials. The main drawbacks are that there is not a unique solution (the final basis may depend on the ordering of the initial system, or on the particular implementation [19], refining algorithms [20, 21], etc.) and that one has to find an ad hoc solution for every type of aperture [12-17]. These, in turn, hinder the physical interpretation of the associated expansion coefficients (especially for the higher orders due to a cumulative effect associated to the G-S method). In addition, the G-S method is especially well-suited for numerical implementation, which means even further optimization for specific parameters of the apertures (eccentricity or orientation of the ellipse, radius of the central obscuration, etc.). Somewhat more general analytical expressions can be obtained too using a non-recursive method [22], but the computational cost may dramatically increase with the order of the polynomial, which may become an effective limitation [11].

In this context, our goal was to develop a general framework able to provide a common formulation under a unique criterion and providing a unique general solution for most of the usual optical apertures. Our approach is based in finding the mapping that transforms the unit circle into the desired aperture geometry. That is finding the diffeomorphism (i.e. a mapping that has a differentiable inverse mapping) [23] that transforms the unit circle into the connected set within the plane that represents the optical aperture. This mapping means warping (and rotating within the plane) the input basis functions so that they fit into the new aperture geometry. On the contrary that ad hoc solutions, that warping permits not only unicity, but also a high level of invariance of the mathematical properties and physical meaning of the basis functions (tilt, defocus, astigmatism, coma, etc.) and hence a natural generalization of the aberration theory. One of the simplest possible mappings consists of the affine transformation (composed of scaling along $x$ and $y$ and rotation) which maps the circle into the ellipse. Here the resulting basis functions are linear combinations of the initial ones (i.e. polynomials), and the associated metrics are proportional (i.e. Euclidean). Other mappings, in particular those transforming the circle into (circular or elliptical) annuli, are non-linear and thus, in general, the warped basis functions are not polynomials and the associated metric may not be Euclidean.

Here we will consider the angular sector of an elliptical annulus with arbitrary orientation as our most general case of mapping of the circle, since other geometries such as circles, ellipses, annuli, sectors, etc. correspond to particular values of the parameters of the general sector. Square, rectangular, hexagonal, etc. geometries are not considered in the present study.

## II. THEORETICAL BASIS

The generalization of unit circle polynomials (or, in general, any complete and orthogonal set of basis functions in the circle) to deformations or partitions of the disc can be achieved by applying a diffeomorphism of the unit circle (or disc) *D* into a connected set *M* within the plane (see Figure 1):

$$\varphi : D \to M \subset R^2, \; \vec{x} := (x, y) = \varphi(u, v) := (x(u, v), y(u, v)) \; . \tag{1}$$

The diffeomorphism is an especially useful transformation in this context, since it is a bijection and its inverse exists and is differentiable as well:

$$\vec{u} := (u, v) = \varphi^{-1}(x, y) := (u(x, y), v(x, y)) \, , \tag{2}$$

and the Jacobian of the inverse transformation $\varphi^{-1}$ will be $J(x, y) := \left( \frac{\partial \vec{u}}{\partial \vec{x}} \right)$. Now, let us choose a complete set of basis functions $Z_j$ (for example Zernike circle polynomials) orthonormal on *D* with metric $dudv$ . These functions are orthonormal also under the change of variables $(u, v) \to (x, y) = \varphi(u, v)$; and taking into account that $dudv = |J(x, y)| \, dxdy$, then we have:

$$\delta_{i,j} = \frac{1}{\pi} \int\int_D Z_i(u, v) Z_j(u, v) dudv = \frac{1}{\pi} \int\int_M Z_i(\varphi^{-1}(x, y)) Z_j(\varphi^{-1}(x, y)) |J(x, y)| \, dxdy \, , \tag{3}$$

where π is the area of the unit circle *D*. This means that the new functions resulting from this change of variables

$$K_j(x, y) := Z_j(\varphi^{-1}(x, y)) \, , \tag{4}$$

are orthonormal on $M$ with metric $|J(x, y)| \, dxdy$ . We can obtain a further generalization by considering the product of these functions with a continuous function $Q(x, y)$ so that:

$$\bar{K}_j(x, y) := Q(x, y) Z_j(\varphi^{-1}(x, y)) \, . \tag{5}$$

The resulting functions are orthonormal on $M$ with metric $Q^{-2}(x, y) |J(x, y)| \, dxdy$. In particular, we can take the trivial case $Q(x, y) = 1$, so that $\bar{K}_j(x, y) = K_j(x, y)$ are orthonormal functions on $M$ with metric $|J(x, y)| \, dxdy$ . Another interesting particular case is when we take $Q(x, y) = \sqrt{|J(x, y)|}$ , in shuch a way that the resulting functions $\bar{K}_j(x, y)$ are orthonormal on $M$ but now with the Euclidean metric $dxdy$ .

It is straightforward to show that the set $\{\bar{K}_j(x,y)\}$ forms a complete system in $L^2(M)$ with metric $Q^{-2}(x,y)\left|J(x,y)\right|dxdy$. For any function $\bar{f}(x,y)$ on $L^2(M)$, we can define $f(u,v)$ on $L^2(D)$ (with Euclidean metric $dudv$) as follows:

$$f(u,v) := Q^{-1}(\varphi(u,v))\,\bar{f}(\varphi(u,v)) = \sum_j c_j Z_j(u,v)\,, \tag{6}$$

that is the expansion of $f$ on the basis set $Z_j$, where the coefficients are given by the projections of $f(u,v)$ on the basis functions:

$$c_j := \frac{1}{\pi}\int\int_D f(u,v) Z_j(u,v)dudv\,. \tag{7}$$

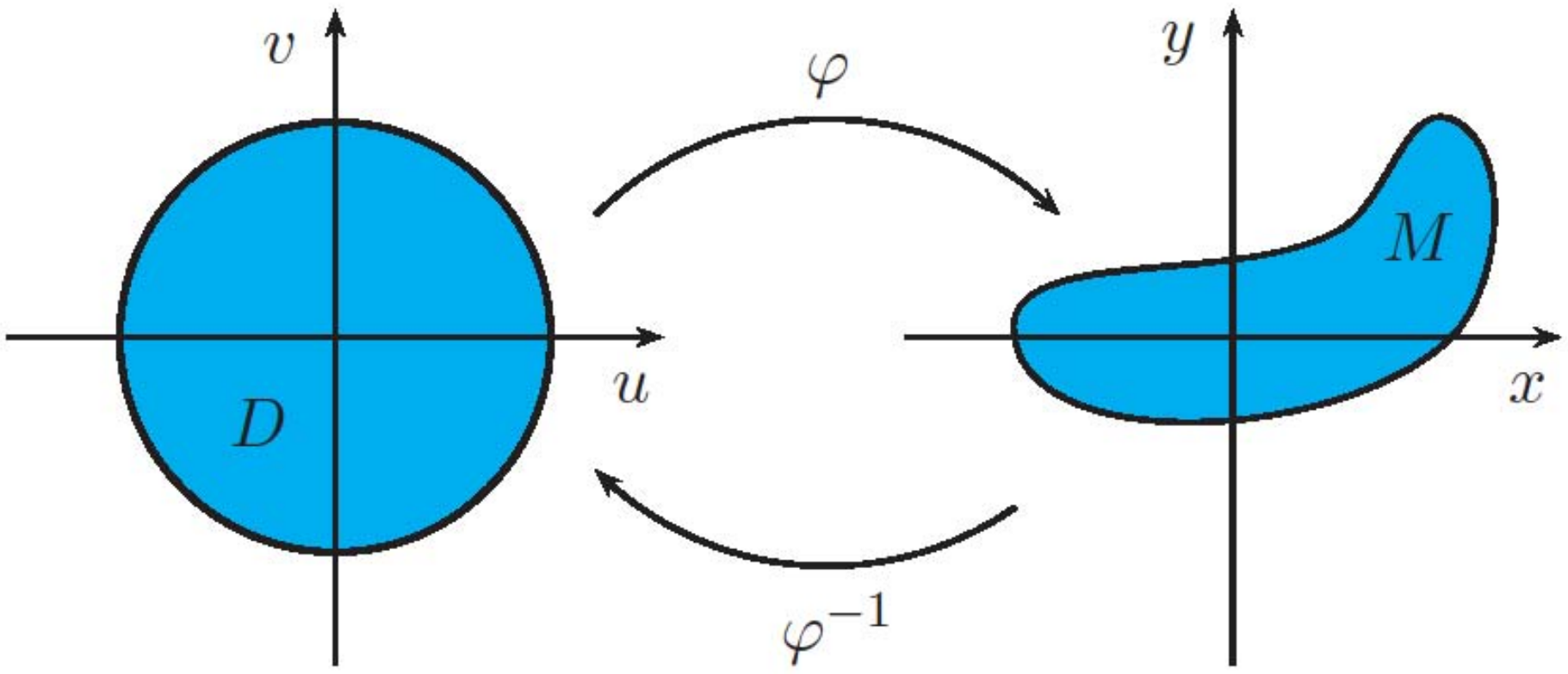

*Fig. 1. Mapping of the unit circle D onto a connected set M through diffeomorphism $\varphi(u,v)$.*

If we now apply the change of coordinates to the definition of the new basis functions, and solve for $Z_j$ we have:

$$Z_j(u,v) = Q^{-1}(\varphi(u,v))\bar{K}_j(\varphi(u,v))\,; \tag{8}$$

and then we find that

$$\bar{f}(x,y) := \sum_j c_j \bar{K}_j(x,y)\,, \tag{9}$$

where the coeffients are also obtained through projection onto the set of basis functions on $M$:

$$c_j := \frac{1}{\pi}\int\int_M \bar{f}(x,y)\bar{K}_j(x,y)Q^{-2}(x,y)\left|J(x,y)\right|dxdy\,. \tag{10}$$

Note that the metric will be $dxdy$ only when $Q(x,y)=\sqrt{|J(x,y)|}$ . The above formulation assumes Cartesian coordinates, but equivalent results are obtained in polar coordinates.

We want to remark that the choice of $Q(x,y)$ has important consequences. In practrice, there are two alternative ways to obtain the expansion coeffients $c_j$, either computing the inner products (projections) with the basis functions, as in Eq. 10, or through least squares fit, which is especially useful for sampled functions [24]. In the last case, the effect of the non Euclidean metric ( $Q(x,y)=1$ ) is that we will have a weighted least squares problem. On the one hand, using $Q(x,y)=\sqrt{|J(x,y)|}$ implies a deformation of the basis functions which might potentially alter their physical meaning. In fact some properties such as zero mean or the possibility of computing the variance of the function as the squared sum of the coefficients can be lost in that case. We know that the mean of functions $K_j(x,y)$ , $\mu_j=\frac{1}{\pi}\iint_M K_j(x,y)|J(x,y)|dxdy=0, \quad \forall j\neq 0$ as a consequence of their orthogonality; but the mean of $\bar{K}_j(x,y)$ , $\bar{\mu}_j=\frac{1}{\pi}\iint_M \bar{K}_j(x,y)dxdy=\frac{1}{\pi}\iint_M K_j(x,y)\sqrt{|J(x,y)|}dxdy$ may be different from zero depending on the Jacobian. The variance $\sigma^2$ of any function $f(x,y)=\sum_j c_j K_j(x,y)$ is $\sigma^2=\frac{1}{\pi}\iint_M f(x,y)^2|J(x,y)|dxdy-\left(\frac{1}{\pi}\iint_M f(x,y)|J(x,y)|dxdy\right)^2$ . If we replace $f(x,y)$ by its expansion $\sum_j c_j K_j(x,y)$ we have $\sigma^2=\sum_j c_j^2$ as far as $K_j(x,y)$ have zero mean ( $\mu_j=0, \forall j\neq 0$ ), since the left integral becomes the sum of $c_j^2\delta_{j,j}$ , whereas the right integrals are equal to zero except for $j$ = 0, for which we have $(c_0)^2$. Thus we have $\sigma^2=\sum_j c_j^2-c_0^2=\sum_{j>0} c_j^2$ . If we now consider the expansion on the system $\bar{K}_j(x,y)$ then we arrive to $\sigma^2=\sum_j \bar{c}_j^2-\left(\sum_j \bar{c}_j\bar{\mu}_j\right)^2\neq\sum_{j>0}\bar{c}_j^2$ . As we discuss further below, despite the complexity added by the non-Euclidean metric to the computation of inner products or to the weighted least squares fit, it seems more convenient to set $Q(x,y)=1$ . Nevertheless, when $J(x,y)=Constant$ then $Q(x,y)=\sqrt{|J(x,y)|}=Constant$ is a simple re-normalization factor.

## III. STANDARD PORTIONS OF CIRCLES AND ELLIPSES

In this section we particularize the above general formulation to standard partitions of circles and ellipses (annuli, angular sectors, etc.). To this end we consider the angular sector $G$ of an elliptical annulus with arbitrary orientation as the most general mapping $\varphi$ considered here. As shown in Table I, the geometry of this general sector $G$ is given by 6 parameters, whereas other shapes (annuli, ellipses, etc.) are obtained as particular cases for specific values of these parameters. The mapping from the unit circle (radius $R$ = 1) into the ellipse can be obtained through a linear affine transformation that is the composition of scaling $x$ and y (to obtain semi axes $A$ and $B$ with $B\leq A$) and rotation by angle $\alpha$ (formed by the major axis $A$ with the $x$ axis). The (concentric) elliptical annulus requires another parameter $0 \le h < 1$, that is the proportionality constant between its inner and outer elliptical boundaries, $h = a/A = b/B$ ($a = hA$ and $b = hB$ are the inner semiaxes). Finally the angular sector will be the area inside the angular interval $[\theta_1, \theta_2]$. Thus, our general mapping will be determined by six parameters: *A, B, α, h,* $\theta_1$ and $\theta_2$. This also includes the mapping of the unit circle into another circle with arbitrary radius $R = A \neq 1$.

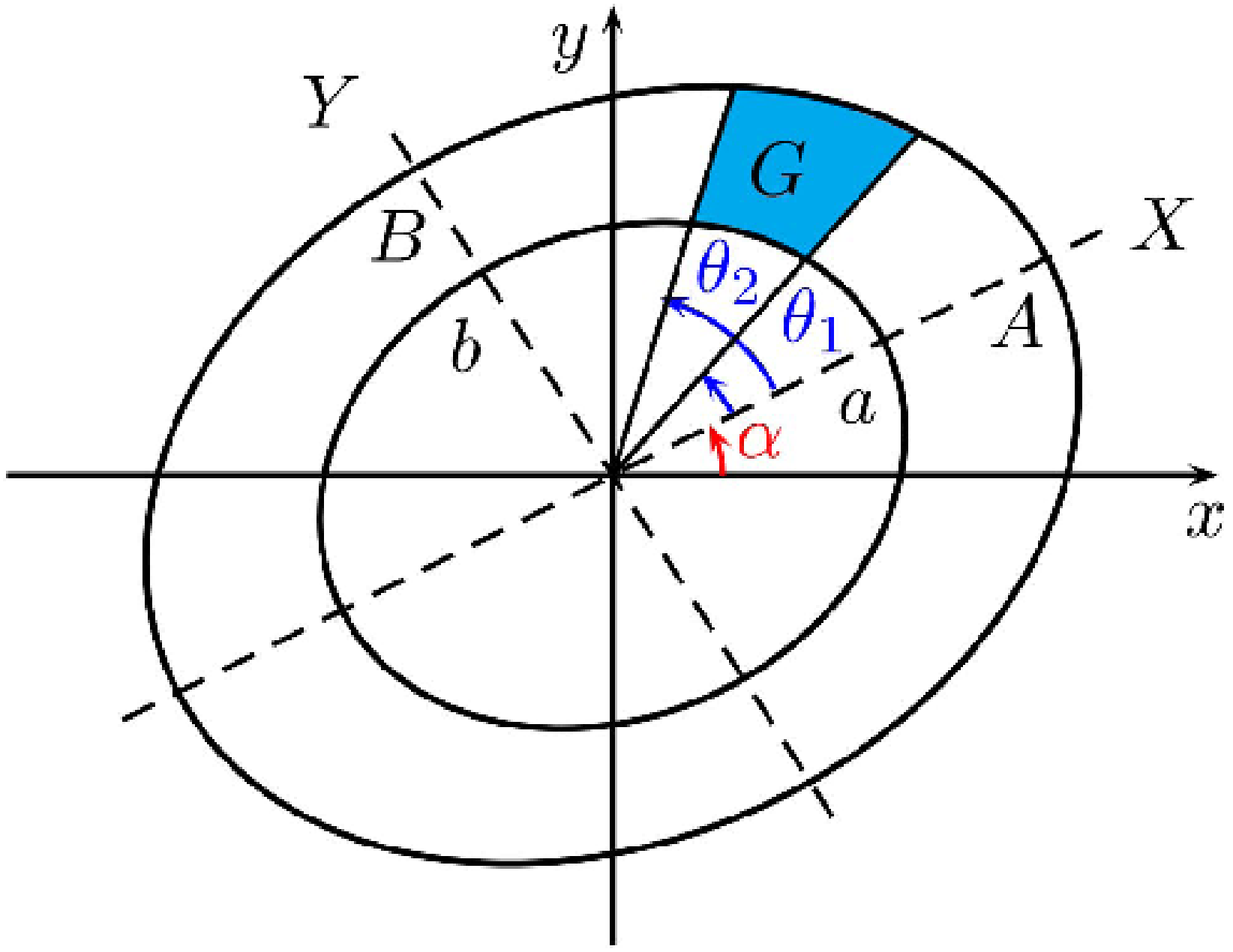

*Fig. 2. General angular sector* $[\theta_1, \theta_2]$ *of an elliptical (semiaxes A, B) annulus (h = a/A= b/B ) with orientation α.*

We can obtain a more compact expression for the mapping $\varphi$ by considering it as the composition of two mappings $\varphi = \varphi_2 \circ \varphi_1$. In the second mapping $\varphi_2$ we apply a rotation by angle $\alpha$ so that:

$$(x, y) = \varphi_2(X, Y) = (X\cos\alpha - Y\sin\alpha, X\sin\alpha + Y\cos\alpha) \quad \text{and} \tag{11a}$$

$$(X, Y) = \varphi_2^{-1}(x, y) = (x\cos\alpha + y\sin\alpha, -x\sin\alpha + y\cos\alpha), \tag{11b}$$

that is a change to the scaled variables $(X, Y) := (AX', BY') := (Ar'\cos\theta', Br'\sin\theta')$, that is $r' := \sqrt{(X/A)^2 + (Y/B)^2}$, $\theta' := \arctan((AY)/(BX))$.

Now we can apply the mapping $\varphi_1$ that transforms the unit circle into the angular ring sector:

$$(X,Y)=\varphi_1(u,v)=(A[h+(1-h)\rho]\cos[\theta_1'+\frac{\theta_2'-\theta_1'}{2\pi}\phi],B[h+(1-h)\rho]\sin[\theta_1'+\frac{\theta_2'-\theta_1'}{2\pi}\phi]) \text{ and } \quad (12a)$$

$$(u,v)=\varphi_1^{-1}(X,Y)=\left(\frac{r'(X,Y)-h}{1-h}\cos\left[\frac{2\pi(\theta'(X,Y)-\theta_1')}{\theta_2'-\theta_1'}\right],\frac{r'(X,Y)-h}{1-h}\sin\left[\frac{2\pi(\theta'(X,Y)-\theta_1')}{\theta_2'-\theta_1'}\right]\right) \quad (12b)$$

where $\rho:=\sqrt{u^2+v^2}$ , $\phi:=\arctan(v/u)$ ; $r'(X,Y)$ and $\theta'(X,Y)$ were defined above. Then, the complete inverse mapping $\varphi^{-1}=\varphi_1^{-1}\circ\varphi_2^{-1}$is:

$$(u,v)=\varphi^{-1}(x,y)=\left(\frac{r'(x,y)-h}{1-h}\cos\left[\frac{2\pi(\theta'(x,y)-\theta_1')}{\theta_2'-\theta_1'}\right],\frac{r'(x,y)-h}{1-h}\sin\left[\frac{2\pi(\theta'(x,y)-\theta_1')}{\theta_2'-\theta_1'}\right]\right), \quad (13)$$

with $r'(x,y)=\sqrt{\left(\frac{x\cos\alpha+y\sin\alpha}{A}\right)^2+\left(\frac{-x\sin\alpha+y\cos\alpha}{B}\right)^2}$ ; $\theta'(x,y)=\arctan\left(\frac{A(-x\sin\alpha+y\cos\alpha)}{B(x\cos\alpha+y\sin\alpha)}\right)$ ; $\theta_1'=\arctan\left[A/_B\tan(\theta_1-\alpha)\right]$; and $\theta_2'=\arctan\left[A/_B\tan(\theta_2-\alpha)\right]$.

As a result of the composition of two transformations, we have an intermediate change of variables$(x,y)\to(X',Y')\to(u,v)$ or in polar coordinates:$(r,\theta)\to(r',\theta')\to(\rho,\phi)$. We can see that$(x,y)=(r\cos\theta,r\sin\theta)$, with $r=\sqrt{x^2+y^2},\theta=\arctan(y/x)$, are now confined within the sector $G$ depicted in Figure 2. The angular interval is $\left[\theta_1:=\alpha+\bar{\theta}_1,\theta_2:=\alpha+\bar{\theta}_2\right]$ where $\bar{\theta}_1,\bar{\theta}_2$ are angles with the X axis. On the one hand, we have $\bar{\theta}=\arctan((B/A)\tan\theta')$ so that $\bar{\theta}_1=\arctan((B/A)\tan\theta_1')$ and $\bar{\theta}_2=\arctan((B/A)\tan\theta_2')$ with $\bar{\theta}_1<\bar{\theta}_2$ and $\bar{\theta}_1,\bar{\theta}_2\in[0,2\pi]$. On the other hand $0\le\rho\le1$ and$0\le\phi\le2\pi$. Then we have:

$$h^2\le\left(\frac{X}{A}\right)^2+\left(\frac{Y}{B}\right)^2=[h+(1-h)\rho]^2=r'^2\le1 \quad \text{and} \quad (14a)$$

$$\theta_1'\le\theta'=\arctan\left(\frac{AY}{BX}\right)=\theta_1'+\frac{\theta_2'-\theta_1'}{2\pi}\phi\le\theta_2'. \quad (14b)$$

In Eq. 14a, we can see that$r'=h+(1-h)\rho$. Since$\rho\in[0,1]$, then $r'\in[h,1]$and Eq. 14b means that $\theta'\in[\theta_1',\theta_2']$ and hence $(x,y)=(r\cos\theta,r\sin\theta)$ take values within the sector $G:=\{(x,y),h\le r'\le1,\theta_1'\le\theta'\le\theta_2'\}$. The Jacobian of this transform is

$$J(x,y)=\frac{2\pi(r'-h)}{ABr'(1-h)^2(\theta_2'-\theta_1')}. \quad (15)$$

It is noteworthy that the Jacobian is not constant only for annuli and annular sectors (see Table I).

A. General system

Now, we can write the new basis functions on G. We will consider $Q(x,y)=1$:

$$\begin{aligned}
&G_j(x,y) := K_j(x,y) = K_j\left(X\cos\alpha - Y\sin\alpha, X\sin\alpha + Y\cos\alpha\right) \\
&= K_j\left(Ar'\cos\theta'\cos\alpha - Br'\sin\theta'\sin\alpha, Ar'\cos\theta'\sin\alpha + Br'\sin\theta'\cos\alpha\right) \\
&:= Z_j\left(\frac{r'-h}{1-h}\cos\left[\frac{2\pi(\theta'-\theta'_1)}{\theta'_2-\theta'_1}\right], \frac{r'-h}{1-h}\sin\left[\frac{2\pi(\theta'-\theta'_1)}{\theta'_2-\theta'_1}\right]\right).
\end{aligned} \tag{16}$$

Functions $G_j$ form a complete orthonormal system on *G*, the general angular sector of the elliptical ring, with metric (differential of surface area) $ds := |J(x,y)|\,dxdy = |J(r\cos\theta, r\sin\theta)|\,rdrd\theta$ . In the Appendix it is shown that functions $G_j$: (1) are orthonormal (i.e. their inner products are Kronecker's deltas); (2) any square-integrable function, defined on *G* can be expressed as a linear combination of functions $G_j$; and (3) when the general sector of an elliptical annulus tends to the unit circle, that is when $A \to 1$ , $B \to 1$ , $\alpha \to 0$ , $h \to 0$ , $\theta_1 \to 0$ and $\theta_2 \to 2\pi$ , then $G_j(x,y) \to Z_j(x,y)$. Table I lists the ranges of the parameters as well as the Jacobian (Eq. 15) for the different particular geometries (annulus, sectors, ellipses, etc.)

B. Zernike polynomials

In what follows we will consider that $Z_j = Z_n^m$ are Zernike circle polynomials (here *j* is a combination of the two indexes: the order of the polynomial *n*, and the angular frequency *m*. Different authors use different ordering, and hence different *j*. Probably the most accepted ordering is the one proposed by Noll [8]). Their expression in polar coordinates $\rho := \sqrt{u^2+v^2}$ , $\phi := \arctan(v/u)$ is:

$$Z_n^m(\rho,\phi) := \begin{cases} N_n^m R_n^{|m|}(\rho)\cos(m\phi), & m \ge 0, \\ -N_n^m R_n^{|m|}(\rho)\sin(m\phi), & m < 0, \end{cases} \quad \text{where} \tag{17a}$$

$$N_n^m = \sqrt{\frac{2(n+1)}{\left(1+\delta_{m0}\right)}} \quad \text{and} \quad R_n^{|m|}(\rho) = \sum_{s=0}^{(n-|m|)/2} \frac{(-1)^s (n-s)!}{s!\left(\frac{n-|m|}{2}-s\right)!\left(\frac{n+|m|}{2}-s\right)!}\rho^{n-2s}. \tag{17b}$$

If we apply the mapping described above, we obtain

$$Z_n^m\left(\frac{r'-h}{1-h}, \frac{2\pi(\theta'-\theta'_1)}{\theta'_2-\theta'_1}\right) = \begin{cases} N_n^m R_n^{|m|}\left(\frac{r'-h}{1-h}\right)\cos\left(m\frac{2\pi(\theta'-\theta'_1)}{\theta'_2-\theta'_1}\right), & m \ge 0, \\ -N_n^m R_n^{|m|}\left(\frac{r'-h}{1-h}\right)\sin\left(m\frac{2\pi(\theta'-\theta'_1)}{\theta'_2-\theta'_1}\right), & m < 0, \end{cases} \tag{18}$$

with $r' = r\sqrt{\left(\frac{\cos(\theta-\alpha)}{A}\right)^2 + \left(\frac{\sin(\theta-\alpha)}{B}\right)^2}$ ; $\theta' = \arctan\left(\frac{A}{B}\tan(\theta-\alpha)\right)$ ; $\theta_1' = \arctan\left[{}^{A}\!/\!{}_{B}\tan(\theta_1-\alpha)\right]$ ; and $\theta_2' = \arctan\left[{}^{A}\!/\!{}_{B}\tan(\theta_2-\alpha)\right]$. Figure 3 shows one example for the particular case of $Z_3^3$ (trefoil). The expression in Cartesian coordinates is given in the Appendix.

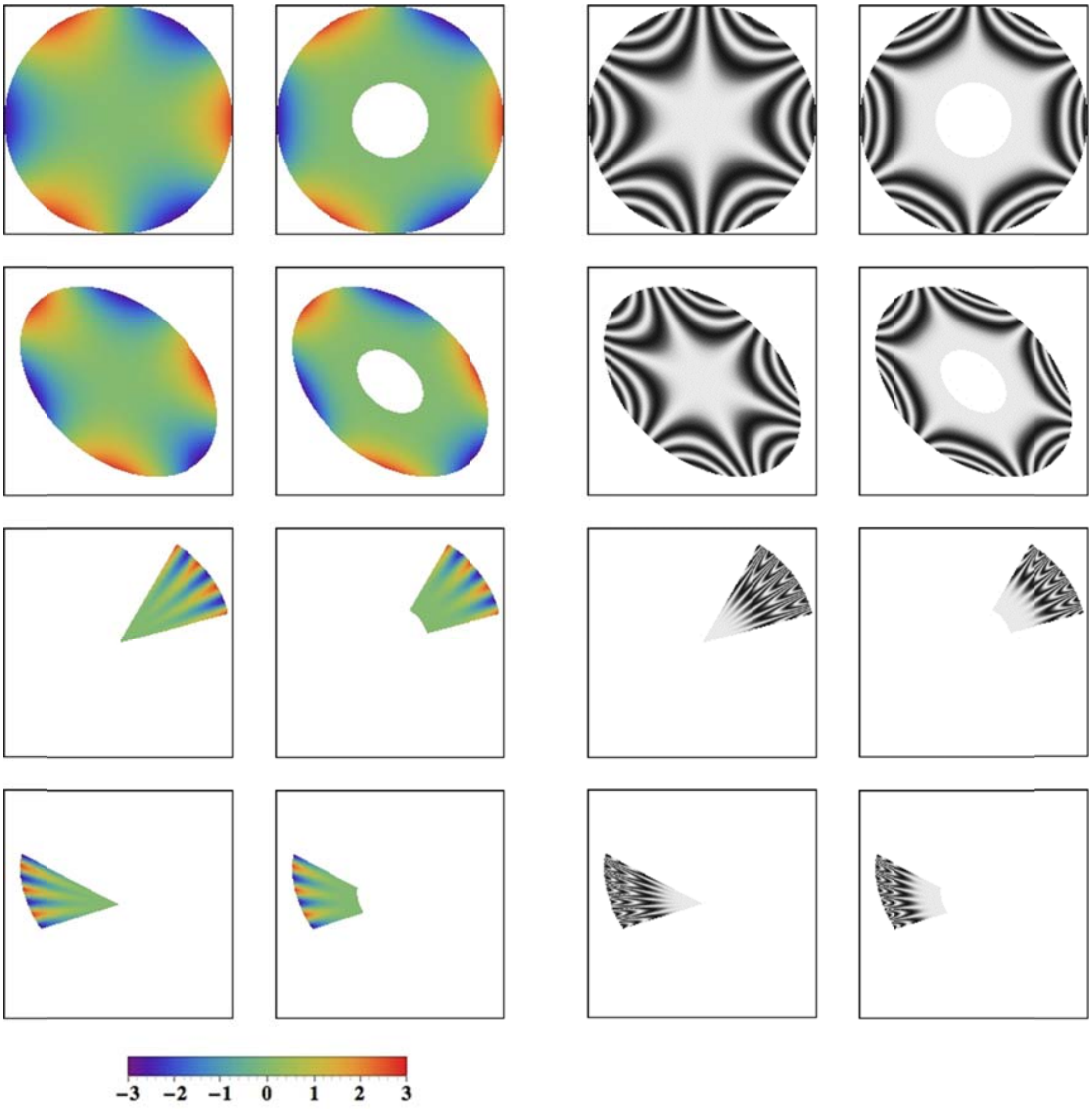

*Fig.3. Basis functions obtained for the particular case of the Zernike polynomial* $Z_3^3$ *(trefoil) corresponding to the different cases listed in Table I. The fringes in the interferograms correspond to 1 wavelength of optical path difference.*

Table I. Range or constant values for the parameters corresponding to the different mappings of the unit circle. All parameters are positive.

| | Basis Functions | Jacobian | *A* | *B* | $\alpha$ | *h* | $\theta_1$ | $\theta_2$ |
|---|---|---|---|---|---|---|---|---|
| **Circle** | $D_j$ | $\frac{1}{A^2}$ | $>0$ | $A$ | 0 | 0 | 0 | $2\pi$ |
| **Annulus** | $O_j$ | $\frac{r-hA}{rA^2(1-h)^2}$ * | $>0$ | $A$ | 0 | $<1$ | 0 | $2\pi$ |
| **Ellipse** | $E_j$ | $\frac{1}{AB}$ | $>0$ | $<A$ | $<\pi$ | 0 | 0 | $2\pi$ |
| **Elliptical Annulus** | $O_j$ | $\frac{r'-h}{r'AB(1-h)^2}$ * | $>0$ | $<A$ | $<\pi$ | $<1$ | 0 | $2\pi$ |
| **Circular sector** | $S_j$ | $\frac{2\pi}{A^2(\theta_2-\theta_1)}$ | $>0$ | $A$ | 0 | 0 | $<\theta_2$ | $\leq 2\pi$ |
| **Elliptical sector** | $S_j$ | $\frac{2\pi}{AB(\theta_2'-\theta_1')}$ | $>0$ | $<A$ | $<\pi$ | 0 | $<\theta_2$ | $\leq 2\pi$ |
| **Annular sector** | $A_j$ | $\frac{2\pi(r-hA)}{rA^2(1-h)^2(\theta_2'-\theta_1')}$ * | $>0$ | $A$ | 0 | $<1$ | $<\theta_2$ | $\leq 2\pi$ |
| **Elliptical annular sector** | $G_j$ | $\frac{2\pi(r'-h)}{r'AB(1-h)^2(\theta_2'-\theta_1')}$ * | $>0$ | $<A$ | $<\pi$ | $<1$ | $<\theta_2$ | $\leq 2\pi$ |

*Non Euclidean metric

## IV. PARTICULAR CASES: ELLIPSES AND ANNULI

In this Section we analyze in detail two cases mostly relevant in optics: ellipses and annuli, departing from the system of Zernike circle polynomials.

### A. Elliptical apertures

As we said above, the mapping from the unit circle into an ellipse is a linear transformation involving three parameters *A*, *B* (it is common to use the eccentricity $e=\sqrt{1-B^2/A^2}$ as a measure of the elongation. This parameter will be used for the explicit expressions of the polynomials listed in Table II), and $\alpha$, so that: $(x,y)=\varphi(u,v)=(Au\cos\alpha-Bv\sin\alpha, Au\sin\alpha+Bv\cos\alpha)$, and its inverse $(u,v)=\varphi^{-1}(x,y)=((x\cos\alpha+y\sin\alpha)/A,(-x\sin\alpha+y\cos\alpha)/B)$. The Jacobian $J(x,y)=1/(AB)$ is constant (i.e. Euclidean metric), so it is possible to use $Q=\sqrt{|J(x,y)|}=1/\sqrt{AB}$ as a simple re-normalization factor. The set of basis functions on the ellipse $E_{A,B,\alpha}:=\{(x,y),(x\cos\alpha+y\sin\alpha)^2/A^2+(-x\sin\alpha+y\cos\alpha)^2/B^2\leq 1\}$ will be:

$$\bar{E}_n^m(x,y) := \frac{1}{\sqrt{AB}} E_n^m(x,y) = \frac{1}{\sqrt{AB}} Z_n^m((x\cos\alpha + y\sin\alpha)/A, (-x\sin\alpha + y\cos\alpha)/B) \qquad (19)$$

For the particular case of Zernike polynomials $Z_n^m(\rho,\phi)$, we only need to simplify Eq. 18 by setting $h=0$; $\theta'_1=0$; and $\theta'_2=2\pi$ to obtain the system $\bar{E}_n^m(r',\theta')$:

$$\bar{E}_n^m(r',\theta') = \begin{cases} \frac{1}{\sqrt{AB}} N_n^m R_n^{|m|}(r')\cos(m\theta'), & m \geq 0, \\ -\frac{1}{\sqrt{AB}} N_n^m R_n^{|m|}(r')\sin(m\theta'), & m < 0, \end{cases} \qquad (20)$$

were $N_n^m$ and $R_n^{|m|}$ were given in Eq.17b and the variables $(r',\theta')$ were defined right after Eq. 18.

Since this is a particular case of *G*, these functions (1) are orthonormal; (2) any square-integrable function, defined on *E* can be expressed as a linear combination of functions $E_n^m$; and (3) when the ellipse tends to the unit circle, that is when $A \to 1$, $B \to 1$ and $\alpha \to 0$, then $E_n^m(x,y) \to Z_n^m(x,y)$. The specific orthogonal elliptical polynomials are listed in Table II up to order 4. Several representative examples, corresponding to various Zernike wavefront aberrations: tilt, defocus, astigmatism, coma, trefoil and spherical aberration are represented in Figure 4 for the particular case of $\alpha$ = 137.5º and *e* = 0.74. This would correspond to the effective pupil for an off-axis (skew) wavefront passing through a circular aperture at field angle 32.4º and azimuth 54.3º (Only positive values of *m* are shown since *m* < 0 are rotated versions of the same aberration modes).

## B. Annular apertures

The mapping of the unit circle into the annulus $\varphi: D \to O$, with *a* and *A* (with *a = hA)* being the radii of the inner and outer circular boundaries respectively, can be expressed as:

$$(x,y) = \varphi(u,v) = (A[h+(1-h)\rho]\cos\phi, A[h+(1-h)\rho]\sin\phi) \qquad (21)$$

and its inverse:

$$(u,v) = \varphi^{-1}(x,y) = \left(\frac{r-hA}{A(1-h)}\cos\theta, \frac{r-hA}{A(1-h)}\sin\theta\right). \qquad (22)$$

With this mapping $r = A\left[h+(1-h)\rho\right]$; since $\rho \in [0,1]$, then $r \in [hA, A]$. This means that the Jacobian $J(x,y) = \frac{r-hA}{rA^2(1-h)^2}$ is positive, but it is not constant, and hence the metric is not Euclidean (unless we set $Q = \sqrt{|J(x,y)|} = \frac{\sqrt{r-hA}}{\sqrt{r}A(1-h)}$). The functions:

$$O_n^m(r,\theta) := Z_n^m\left(\frac{r-hA}{A(1-h)}\cos\theta, \frac{r-hA}{A(1-h)}\sin\theta\right) \text{ or} \qquad (23a)$$

Table II. Expressions of the orthogonal elliptical polynomials up to order n = 4 where A is the major semi axis; $e=\sqrt{1-B^2/A^2}$ is the eccentricity; and $\alpha$ is the orientation of the ellipse.

**Elliptical polynomials**

$$E_0^0 = 1$$

$$E_1^1 = \frac{2(x\cos\alpha + y\sin\alpha)}{A}$$

$$E_1^{-1} = \frac{2(y\cos\alpha - x\sin\alpha)}{A\sqrt{1-e^2}}$$

$$E_2^0 = \frac{\sqrt{3}\left(-A^2\left(e^2-1\right)+\left(e^2-2\right)\left(x^2+y^2\right)+e^2(2xy\sin 2\alpha+\cos 2\alpha(x^2-y^2))\right)}{A^2\left(e^2-1\right)}$$

$$E_2^{-2} = \frac{\sqrt{6}\left(\sin 2\alpha\left(y^2-x^2\right)+2xy\cos 2\alpha\right)}{A^2\sqrt{1-e^2}}$$

$$E_2^2 = \frac{\sqrt{3}\left(e^2\left(x^2+y^2\right)+\left(e^2-2\right)(2xy\sin 2\alpha+\cos 2\alpha(x^2-y^2))\right)}{\sqrt{2}A^2\left(e^2-1\right)}$$

$$E_3^{-1} = \sqrt{2}\frac{\sqrt{1-e^2}\,(y\cos\alpha - x\sin\alpha)\left(4A^2\left(e^2-1\right)-3\left(e^2-2\right)\left(x^2+y^2\right)-3e^2(2xy\sin 2\alpha+\cos 2\alpha(x^2-y^2))\right)}{A^3\left(e^2-1\right)^2}$$

$$E_3^1 = \frac{\sqrt{2}(x\cos\alpha + y\sin\alpha)\left(-4A^2\left(e^2-1\right)+3\left(e^2-2\right)\left(x^2+y^2\right)+3e^2(2xy\sin 2\alpha+\cos 2\alpha(x^2-y^2))\right)}{A^3\left(e^2-1\right)}$$

$$E_3^{-3} = \frac{\sqrt{2}(y\cos\alpha - x\sin\alpha)\left(-\left(3e^2-2\right)\left(x^2+y^2\right)-\left(3e^2-4\right)(2xy\sin 2\alpha+\cos 2\alpha(x^2-y^2))\right)}{A^3\left(1-e^2\right)^{3/2}}$$

$$E_3^3 = \frac{\sqrt{2}(x\cos\alpha + y\sin\alpha)\left(\left(e^2+2\right)\left(x^2+y^2\right)+(e^2-4)(2xy\sin 2\alpha+\cos 2\alpha(x^2-y^2))\right)}{A^3\left(e^2-1\right)}$$

$$E_4^0 = \frac{\sqrt{5}}{4A^4\left(e^2-1\right)^2}\left(\begin{array}{l} 4A^4\left(e^2-1\right)^2+3e^2\left(\begin{array}{l} 8xy\sin 2\alpha\left(\left(e^2-2\right)\left(x^2+y^2\right)-A^2\left(e^2-1\right)\right) \\ +4\cos 2\alpha(x^2-y^2)\left(\left(e^2-2\right)\left(x^2+y^2\right)-A^2\left(e^2-1\right)\right) \\ +e^2\cos 4\alpha\left(x^4-6x^2y^2+y^4\right)+4e^2xy\sin 4\alpha(x^2-y^2) \end{array}\right) \\ -12A^2\left(e^4-3e^2+2\right)\left(x^2+y^2\right)+3\left(3e^4-8e^2+8\right)\left(x^2+y^2\right)^2 \end{array}\right)$$

$$E_4^2 = \frac{\sqrt{5}}{\sqrt{2}A^4\left(e^2-1\right)^2}\left(e^2\left(x^2+y^2\right)+\left(e^2-2\right)(2xy\sin 2\alpha+\cos 2\alpha(x^2-y^2))\right)$$
$$\times\left(-3A^2\left(e^2-1\right)+2\left(e^2-2\right)\left(x^2+y^2\right)+2e^2(2xy\sin 2\alpha+\cos 2\alpha(x^2-y^2))\right)$$

$$E_4^{-2} = -\frac{\sqrt{10}(\sin 2\alpha(x^2-y^2)-2xy\cos 2\alpha)\left(3A^2\left(e^2-1\right)-2\left(e^2-2\right)\left(x^2+y^2\right)-2e^2(2xy\sin 2\alpha+\cos 2\alpha(x^2-y^2))\right)}{A^4\left(1-e^2\right)^{3/2}}$$

$$E_4^4 = \frac{\sqrt{5}}{4\sqrt{2}A^4\left(e^2-1\right)^2}\left(\begin{array}{l} 3e^4\left(x^2+y^2\right)^2+4\left(e^2-2\right)e^2\cos 2\alpha\left(x^4-y^4\right)+8\left(e^2-2\right)e^2xy\sin 2\alpha\left(x^2+y^2\right) \\ +\left(e^4-8e^2+8\right)\cos 4\alpha\left(x^4-6x^2y^2+y^4\right)+4\left(e^4-8e^2+8\right)xy\sin 4\alpha(x^2-y^2) \end{array}\right)$$

$$E_4^{-4} = \frac{\sqrt{10}(\sin 2\alpha(x^2-y^2)-2xy\cos 2\alpha)\left(e^2\left(x^2+y^2\right)+\left(e^2-2\right)(2xy\sin 2\alpha+\cos 2\alpha(x^2-y^2))\right)}{A^4\left(1-e^2\right)^{3/2}}$$

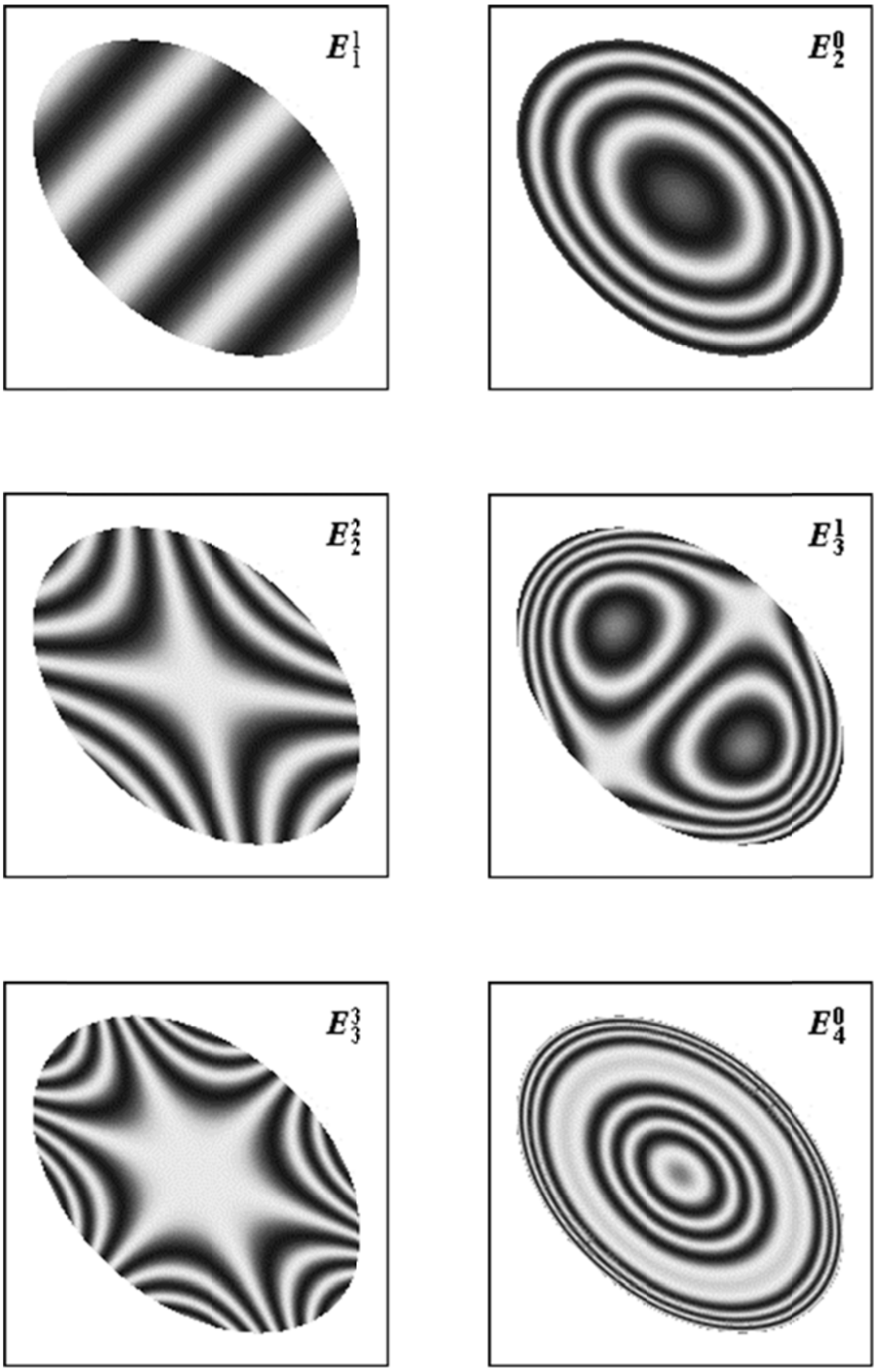

*Fig.4. Orthogonal elliptical polynomials corresponding to Zernike wavefront aberrations: tilt* $E_1^1$*, defocus* $E_2^0$*, astigmatism* $E_2^2$*, coma* $E_3^1$*, trefoil* $E_3^3$ *and spherical aberration* $E_4^0$*. The fringes in the interferograms correspond to 1 wavelength of optical path difference.*

$$\bar{O}_n^m(r,\theta) := \frac{\sqrt{r-hA}}{\sqrt{r}A(1-h)} Z_n^m\left(\frac{r-hA}{A(1-h)}\cos\theta, \frac{r-hA}{A(1-h)}\sin\theta\right) \tag{23b}$$

form complete orthonormal systems on the annulus $O := \{(x,y), a = hA \le r \le A, 0 \le \theta \le 2\pi\}$ with metrics $ds := \frac{r-hA}{rA^2(1-h)^2}dxdy = \frac{r-hA}{A^2(1-h)^2}drd\theta$ for system $\{O_n^m\}$ or $ds := dxdy = rdrd\theta$ for system $\{\bar{O}_n^m\}$ . The general expression in polar coordinates for $\{O_n^m\}$ results:

$$O_n^m(r,\theta) = \begin{cases} \sqrt{\frac{2(n+1)}{(1+\delta_{m0})}} \sum_{s=0}^{(n-|m|)/2} \frac{(-1)^s (n-s)!}{s!\left(\frac{n+|m|}{2}-s\right)!\left(\frac{n-|m|}{2}-s\right)!} \left(\frac{r-hA}{A(1-h)}\right)^{n-2s} \cos(m\theta), & m \ge 0, \\ -\sqrt{\frac{2(n+1)}{(1+\delta_{m0})}} \sum_{s=0}^{(n-|m|)/2} \frac{(-1)^s (n-s)!}{s!\left(\frac{n+|m|}{2}-s\right)!\left(\frac{n-|m|}{2}-s\right)!} \left(\frac{r-hA}{A(1-h)}\right)^{n-2s} \sin(m\theta), & m < 0. \end{cases} \tag{24}$$

Again, this system is complete, orthonormal and tends to $Z_n^m$ when $h \to 0$ and $A \to 1$. The specific annular functions (polynomial quotients) are listed in Table III up to order 4, and representative examples are shown in Figure 5 for $h$ = 0.33 which is close to that of the central obscuration of the Hubble telescope.

## V. DISCUSSION AND CONCLUSIONS

We presented a general method to obtain complete orthonormal systems in any connected set $M$. It consists of applying a diffeomorphism to a given (complete and orthonormal) system on the unit circle to transform the circle basis functions into the new system on the desired geometry. This type of mapping is a bijection, so that it is invertible. In addition, its inverse is differentiable, so it has an associated Jacobian. All these properties are essential for establishing a rigorous and robust theoretical framework. We then particularized the method to circles, ellipses and their standard portions: annuli and angular sectors, etc. The general mapping of the unit circle into an angular sector of an elliptical ring can be obtained by the composition of four mappings: for the annuli: $r \to (r-h)/(1-h)$ where $h$ and 1 are the radii of the inner and outer (unit disk) boundaries of the annulus. Similarly for the angular sectors: $\theta \to 2\pi(\theta-\theta_1)/(\theta_2-\theta_1)$. The scaling: $x \to Ax$ and $y \to By$ transforms the unit circle into a horizontal ellipse. Finally the in-plane rotation by angle $\alpha$ permits to have an arbitrary orientation. The eight geometries, listed in Table I, are obtained by composing one, two, three or four of these transformations. The associated Jacobians are constant except for the radial mapping $r \to (r-h)/(1-h)$, since the Jacobian to pass from Cartesian to polar coordinates is not constant. Thus, the simplicity of this framework, based on simple changes of variables is another interesting aspect (even though the composition of up to four of these changes may yield to long expressions).

These systems are especially important in the wave theory of aberrations and optical image formation [7, 5]. As we said in the Introduction, there is not a complete agreement on a unified theory yet. Nevertheless, the good mathematical properties ZPs are making that they are becoming

Table III. Expressions of the orthogonal annular basis functions (polynomial quotients) up to order n = 4. *A* and $a = hA$ are the outer and inner radii.

Annular polynomials

$$O_0^0 = 1$$

$$O_1^1 = \frac{2x}{A(1-h)}\left(1 - \frac{hA}{\sqrt{x^2+y^2}}\right)$$

$$O_1^{-1} = \frac{2y}{A(h-1)}\left(\frac{hA}{\sqrt{x^2+y^2}} - 1\right)$$

$$O_2^2 = \frac{\sqrt{3}\left(A^2(h(h+2)-1) - 4hA\sqrt{x^2+y^2} + 2\left(x^2+y^2\right)\right)}{A^2(h-1)^2}$$

$$O_2^{-2} = \frac{2\sqrt{6}xy\left(A^2h^2 - 2hA\sqrt{x^2+y^2} + x^2+y^2\right)}{A^2(h-1)^2\left(x^2+y^2\right)}$$

$$O_2^2 = \frac{\sqrt{6}(x^2-y^2)\left(A^2h^2 - 2hA\sqrt{x^2+y^2} + x^2+y^2\right)}{A^2(h-1)^2\left(x^2+y^2\right)}$$

$$O_3^{-1} = \frac{2\sqrt{2}y\left(-\sqrt{x^2+y^2}\left(A^2(h(7h+4)-2) + 3\left(x^2+y^2\right)\right) + h(h(h+4)-2)A^3 + 9hA\left(x^2+y^2\right)\right)}{A^3(h-1)^3\sqrt{x^2+y^2}}$$

$$O_3^1 = \frac{2\sqrt{2}x\left(-\sqrt{x^2+y^2}\left(A^2(h(7h+4)-2) + 3\left(x^2+y^2\right)\right) + h(h(h+4)-2)A^3 + 9hA\left(x^2+y^2\right)\right)}{A^3(h-1)^3\sqrt{x^2+y^2}}$$

$$O_3^{-3} = \frac{2\sqrt{2}y\left(y^2-3x^2\right)\left(\sqrt{x^2+y^2}\left(3A^2h^2 + x^2+y^2\right) - h^3A^3 - 3hA\left(x^2+y^2\right)\right)}{A^3(h-1)^3\left(x^2+y^2\right)^{3/2}}$$

$$O_3^{-3} = -\frac{2\sqrt{2}x\left(x^2-3y^2\right)\left(\sqrt{x^2+y^2}\left(3A^2h^2 + x^2+y^2\right) - h^3A^3 - 3hA\left(x^2+y^2\right)\right)}{A^3(h-1)^3\left(x^2+y^2\right)^{3/2}}$$

$$O_4^0 = \frac{\sqrt{5}}{A^4(h-1)^4}\left(\begin{array}{l} A^4\left(h^4+8h^3-4h+1\right) - 12hA\sqrt{x^2+y^2}\left(A^2(h(h+2)-1) + 2\left(x^2+y^2\right)\right) \\ +6A^2(h(5h+2)-1)\left(x^2+y^2\right) + 6\left(x^2+y^2\right)^2 \end{array}\right)$$

$$O_4^2 = \frac{\sqrt{10}(x^2-y^2)}{A^4(h-1)^4\left(x^2+y^2\right)}\left(\begin{array}{l} A^4h^2(h(h+6)-3) - 2hA\sqrt{x^2+y^2}\left(A^2(h(5h+6)-3) + 8\left(x^2+y^2\right)\right) \\ +3A^2(h(7h+2)-1)\left(x^2+y^2\right) + 4\left(x^2+y^2\right)^2 \end{array}\right)$$

$$O_4^{-2} = \frac{2\sqrt{10}xy}{A^4(h-1)^4\left(x^2+y^2\right)}\left(\begin{array}{l} A^4h^2(h(h+6)-3) - 2hA\sqrt{x^2+y^2}\left(A^2(h(5h+6)-3) + 8\left(x^2+y^2\right)\right) \\ +3A^2(h(7h+2)-1)\left(x^2+y^2\right) + 4\left(x^2+y^2\right)^2 \end{array}\right)$$

$$O_4^4 = \frac{\sqrt{10}\left(x^4-6x^2y^2+y^4\right)\left(A^4h^4 - 4hA\sqrt{x^2+y^2}\left(A^2h^2+x^2+y^2\right) + 6A^2h^2\left(x^2+y^2\right) + \left(x^2+y^2\right)^2\right)}{A^4(h-1)^4\left(x^2+y^2\right)^2}$$

$$O_4^{-4} = \frac{4\sqrt{10}xy(x^2-y^2)\left(A^4h^4 - 4hA\sqrt{x^2+y^2}\left(A^2h^2+x^2+y^2\right) + 6A^2h^2\left(x^2+y^2\right) + \left(x^2+y^2\right)^2\right)}{A^4(h-1)^4\left(x^2+y^2\right)^2}$$

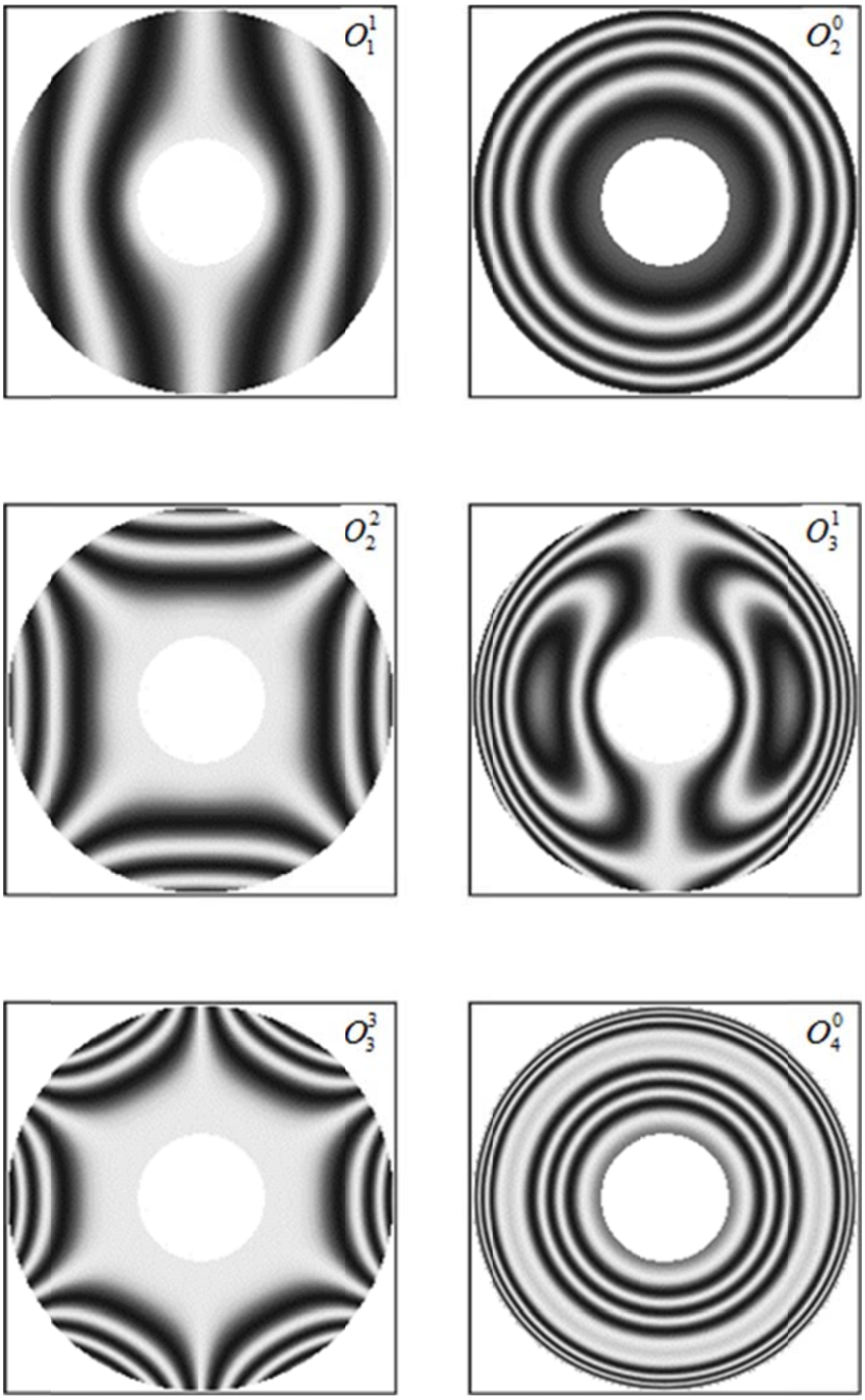

*Fig.5. Orthogonal annular polynomials corresponding to Zernike wavefront aberrations: tilt* $O_1^1$*, defocus* $O_2^0$*, astigmatism* $O_2^2$*, coma* $O_3^1$*, trefoil* $O_3^3$ *and spherical aberration* $O_4^0$*. The fringes in the interferograms correspond to 1 wavelength of optical path difference.*

a standard among the scientific community, but even more in technologies embedded in many industrial tools (optical design), devices (surface metrology, optical testing) and even in clinical apparatus in ophthalmology (corneal topographers, ocular aberrometers, etc.) In addition to form a complete orthonormal system they have compact expressions both in polar and Cartesian coordinates, and they also show an interesting list of additional properties. The main drawbacks come from the fact that they represent a highly convenient but arbitrary choice, and from some difficulties for unification with the Seidel (third order and further extensions to 5th and higher orders) theory of aberrations. The wave aberration theory based on orthogonal systems, such as Zernike polynomials, is superior to the Seidel theory for nearly diffraction-limited optical system because the image quality, as measured by the Strehl ratio (peak of the point spread function normalized to that of the Airy disk) [5] can be predicted from the wavefront variance, and hence from the expansion coefficients as: $Sr \approx 1-k^2\sigma^2 = 1-k^2\sum_{j>0} {c_j}^2$ (where $k = 2\pi/\lambda$ is the wave number). However this approximation is valid only when the aberrations are small, i.e. when $\sigma$ is not much greater than $\lambda/14$ (Maréchal criterion for diffraction-limited optical quality) [5]. Thus, the wavefront variance $\sigma^2$, or the root mean square (RMS) wavefront error $\sigma$, are good metrics for optical image only when $\sigma << \lambda$. Thus the quadratic sum of the wavefront coefficients $\sigma^2$, or the RMS $\sigma$, are good metrics for the wavefront quality, but in aberrated systems they are not good metrics for predicting image quality.

Nevertheless, the need for a unified theoretical framework for describing wavefronts (optical aberrations) and optical surfaces, in terms of complete orthogonal systems, is patent. Even in the context of Zernike polynomials one can find that different authors use a variety of units (wavelengths, micrometers, etc.), normalizations (use of $N_n^m$ or not), ordering (since they have two indexes *n, m* there are many different possibilities to define a single index *j* which were used by different authors), or even opposite sign conventions. Among all these possibilities the ordering proposed by Noll [8] seems widely accepted, but the Optical Society of America adopted a different ordering for reporting aberrations in the human eye (ANSI standard Z80.28). This disparity of criteria existing even for the Zernike circle polynomials may explode when extending the Zernike theory to other optical apertures, such ellipses, annuli, sectors, etc. In fact there is a large list of publications with different polynomials which are *ad hoc* solutions for every type of optical apertures [11-18]. Most of these polynomials are obtained through G-S orthonormalization. A relevant improvement over the classical G-S method can be obtained by a non-recursive algorithm [19] which shows better numerical stability. It consists of inverting the Gram matrix, but considering only up to a given order *n*, and using the Cholesky decomposition. Additional numerical stability can be obtained through re-orthogonalization [20] or by an iterative process [21] up to machine precision accuracy. This non-recursive G-S algorithm provided probably the best results published in the literature [18]. The main limitation is that the symbolic implementation of these methods requires heavy loads of computer memory [11]. We want to remark that the systems of orthogonal polynomials obtained with these improved G-S methods are perfectly valid for ad hoc applications. However, unicity is not guaranteed as the result depends on the initial ordering (index *j*), or on the particular algorithm. In addition, the physical meaning of each polynomial could change after orthogonalization, especially for higher order polynomials, due to the cumulative effect inherent to the G-S method. As a result, the basis functions obtained for different apertures may have different properties and physical meaning, which seems far from the goal of a unified theory.

We believe that the mapping method proposed here, implemented as a change of variables, overcomes most of these difficulties and drawbacks, and provides a common framework, especially well-suited for a unified theory of aberrations. In addition to the theoretical relevance, the unified formulation may be the starting point for developing standards for academic, industrial or even clinical (ophthalmic optics) use.

The two particular cases of Section IV are especially relevant. In the first case, the importance is clear if we realize even the circular aperture (most common in optics) becomes elliptical off-axis. The size ($A$) eccentricity ($e$) and orientation ($\alpha$) of the ellipse changes continuously with field angle and azimuth. These changes can be large in wide angle optical systems, and especially important in vision (both biological and artificial), where the field of view can be of the order of 180º or even more. Thus, the need for having a unified method to describe wavefronts corresponding to different field angles seems clear. The proposed change of variables (mapping) is a simple affine transformation from the circle to the ellipse, to adapt the modes (polynomials) to the scales ($A$ and $B$) and orientation of the effective aperture, which changes as we move from the optical axis to a peripheral field angle. This affine transformation (scaling and rotation) is general for virtually any type of aperture (annuli, polygons, etc.) while it keeps a reasonable level of invariance, and hence a similar physical meaning of the expansion coefficients. This mapping should be applied for representing off-axis wavefronts, especially when the field of view is significantly wide.

The case of annular apertures is relevant not only for being a typical aperture in telescopes, but also due to important differences associated to its particular topology. Other apertures, such as circles, ellipses, or even angular sectors are simply connected sets, whereas the central obscuration of annuli makes that they are connected (but not simply) sets. This has important consequences. The deformation (mapping) of the circle necessary to arrive at the annulus (intuitively this would be a sort of stretching of an infinitesimal central hole to form the finite obscuration) is not uniform, as it depends on the distance to the center ($\rho$). This lack of uniformity makes that the original polynomials become polynomial quotients (see Table III) and that the Jacobian is not a simple constant normalization factor. Then here we have a non-Euclidean metric and hence the original plane element of area becomes curved now. That curvature becomes patent if we compare the upper left panels of Figures 4 and 5 which represent $E_1^1$ (ellipse) and $O_1^1$ (annulus) respectively. In $E_1^1$ the mapping is linear and hence we can see tilted (rotated) but straight fringes, whereas $O_1^1$ shows vertical but curved fringes. As we said in Section II, it is possible to choose $Q(x,y)=\sqrt{|J(x,y)|}$ but this does not eliminate the curvature of the fringes. This choice does not seem compatible with a unified framework as it can potentially change the properties of the basis functions and the physical meaning of the associated coefficients.

## Acknowledgments

This research was supported by the Spanish Ministry of Economía y Competitividad and the European Union, grant FIS2011-22496, and by the Government of Aragón, research group E99.

## APPENDIX

We can see that the functions $G_j$ (defined on a general elliptical annular sector in Section III) are:

A. Orthonormal

The inner product between two functions in the system $\{G_j(x,y)\}$ is

$$\frac{1}{\pi}\iint_G G_i(x,y)G_j(x,y)\left|J(x,y)\right|dxdy = \frac{1}{\pi}\iint_{G'} K_i(X\cos\alpha - Y\sin\alpha, X\sin\alpha + Y\cos\alpha)$$
$$\times K_j(X\cos\alpha - Y\sin\alpha, X\sin\alpha + Y\cos\alpha)\,|\,J(X\cos\alpha - Y\sin\alpha, X\sin\alpha + Y\cos\alpha)\,|\,dXdY$$

where we applied the intermediate change of variables $(X,Y) = (x\cos\alpha + y\sin\alpha, -x\sin\alpha + y\cos\alpha)$ so that $G'$ is the sector rotated by $-\alpha$ (i.e. canonical orientation). Now we pass to polar coordinates $(r',\theta')$ and include the expression of the Jacobian:

$$= \frac{1}{\pi}\iint_{G'} K_i(Ar'\cos\theta'\cos\alpha - Br'\sin\theta'\sin\alpha, Ar'\cos\theta'\sin\alpha + Br'\sin\theta'\cos\alpha)$$
$$\times K_j(Ar'\cos\theta'\cos\alpha - Br'\sin\theta'\sin\alpha, Ar'\cos\theta'\sin\alpha + Br'\sin\theta'\cos\alpha)\frac{2\pi(r'-h)dr'd\theta'}{(1-h)^2(\theta_2'-\theta_1')}$$

then using the definition of the functions $K_i$

$$= \frac{2}{\theta_2'-\theta_1'}\int_h^1 dr'\int_{\theta_1'}^{\theta_2'} d\theta' \frac{r'-h}{(1-h)^2} Z_i\left(\frac{r'-h}{1-h}\cos\left[\frac{2\pi(\theta'-\theta_1')}{\theta_2'-\theta_1'}\right], \frac{r'-h}{1-h}\sin\left[\frac{2\pi(\theta'-\theta_1')}{\theta_2'-\theta_1'}\right]\right)$$
$$\times Z_j\left(\frac{r'-h}{1-h}\cos\left[\frac{2\pi(\theta'-\theta_1')}{\theta_2'-\theta_1'}\right], \frac{r'-h}{1-h}\sin\left[\frac{2\pi(\theta'-\theta_1')}{\theta_2'-\theta_1'}\right]\right)$$

and finally we apply the change of variables $(r'\to\rho, \theta'\to\phi)$, and pass from polar to Cartesian $(u,v)$ coordinates we obtain:

$$\frac{1}{\pi}\iint_G G_i(r,\theta)G_j(r,\theta)\left|J(x,y)\right|rdrd\theta = \frac{1}{\pi}\int_0^1 \rho d\rho\int_0^{2\pi} d\phi Z_i(\rho,\phi)Z_j(\rho,\phi) = \frac{1}{\pi}\iint_D Z_i(u,v)Z_j(u,v) = \delta_{i,j}$$

B. Complete

For any function $\bar{f}(x,y)\in L^2(G)$, square integrable on the general sector $G$, we can define $f(u,v)$ on $f(u,v)\in L^2(D)$ as:

$$f(u,v) := \bar{f}\left(A[h+(1-h)\rho]\cos\left[\theta_1' + \frac{\theta_2'-\theta_1'}{2\pi}\phi\right], B[h+(1-h)\rho]\sin\left[\theta_1' + \frac{\theta_2'-\theta_1'}{2\pi}\phi\right]\right).$$ Then

$$\overline{f}(x,y)=f\left(\frac{r'-h}{1-h}\cos\left[\frac{2\pi(\theta'-\theta'_1)}{\theta'_2-\theta'_1}\right],\frac{r'-h}{1-h}\sin\left[\frac{2\pi(\theta'-\theta'_1)}{\theta'_2-\theta'_1}\right]\right)$$

$$=\sum c_j Z_j\left(\frac{r'-h}{1-h}\cos\left[\frac{2\pi(\theta'-\theta'_1)}{\theta'_2-\theta'_1}\right],\frac{r'-h}{1-h}\sin\left[\frac{2\pi(\theta'-\theta'_1)}{\theta'_2-\theta'_1}\right]\right),$$

and after Eq. 16: $\overline{f}(x,y)=\sum c_j G_j\left(r\cos\theta,r\sin\theta\right)=\sum c_j G_j\left(x,y\right)$.

C. They tend to $Z_j$ when G tends to D

$$\lim_{\substack{A\to1\\B\to1\\\alpha\to0\\h\to0\\\theta_1\to0\\\theta_2\to2\pi}} G_j\left(x,y\right)=\lim_{\substack{A\to1\\B\to1\\\alpha\to0\\h\to0\\\theta_1\to0\\\theta_2\to2\pi}} Z_j\left(\frac{r'-h}{1-h}\cos\left[\frac{2\pi(\theta'-\theta'_1)}{\theta'_2-\theta'_1}\right],\frac{r'-h}{1-h}\sin\left[\frac{2\pi(\theta'-\theta'_1)}{\theta'_2-\theta'_1}\right]\right)$$

$$=\lim_{\substack{r'\to r\\\theta'\to\theta}} Z_j\left(r'\cos\theta',r'\sin\theta'\right)=Z_j\left(r\cos\theta,r\sin\theta\right)=Z_j\left(x,y\right).$$

D. General expression in Cartesian coordinates

For the particular case of Zernike circle polynomials, the expression in Cartesian coordinates can be obtained by applying the corresponding changes of variables to the equation 13.40 in [25] (variables *x* and *y* were interchanged according to the different convention used for index *m* in [25]) :

$$G_n^m(x,y)=\sqrt{\frac{2(n+1)}{\left(1+\delta_{m0}\right)}}\sum_{i=0}^{q}\sum_{j=0}^{m}\sum_{k=0}^{m-j}(-1)^{i+j}\frac{(n-j)!}{j!(m-j)!(n-m-j)!}$$

$$\times\left(\frac{r'-h}{1-h}\sin\left[\frac{2\pi(\theta'-\theta'_1)}{\theta'_2-\theta'_1}\right]\right)^{2(i+k)+p}\left(\frac{r'-h}{1-h}\cos\left[\frac{2\pi(\theta'-\theta'_1)}{\theta'_2-\theta'_1}\right]\right)^{n-2(i+j+k)-p}$$

where *s* and *q* are:

$$n-2m\le0\begin{cases}n\left(even\right)\to p=0;\ q=-\dfrac{n-2m}{2}\\ n\left(odd\right)\to p=0;\ q=-\dfrac{n-2m+1}{2}\end{cases}$$

$$n-2m>0\begin{cases}n\left(even\right)\to p=1;\ q=-\dfrac{n-2m}{2}\\ n\left(odd\right)\to p=1;\ q=-\dfrac{n-2m-1}{2}\end{cases}$$

with $r'=\sqrt{\left(\frac{x\cos\alpha+y\sin\alpha}{A}\right)^2+\left(\frac{-x\sin\alpha+y\cos\alpha}{B}\right)^2}$ ; $\theta'=\arctan\left(\frac{A(-x\sin\alpha+y\cos\alpha)}{B(x\cos\alpha+y\sin\alpha)}\right)$ ;

$\theta'_1=\arctan\left[{}^{A}\!/_{B}\tan(\theta_1-\alpha)\right]$; and $\theta'_2=\arctan\left[{}^{A}\!/_{B}\tan(\theta_2-\alpha)\right]$.

## REFERENCES

[1 ] F. Zernike, “Beugungstheorie des schneidenverfahrens und seiner verbesserten form, der phasenkontrastmethode”, *Physica (Utrecht)* 1, 689 (1934).

[2] J. Schwiegerling, J. Greivenkamp, and J. Miller, “Representation of videokeratoscopic height data with Zernike polynomials”, J. *Opt. Soc. Am. A* 12, 2105 (1995).

[3] D. R. Iskander, M. J. Collins, and B. Davis, “Optimal modeling of corneal surfaces with Zernike polynomials”, *IEEE Trans. Biomed. Eng*. 48, 87 (2001).

[4] W. Lukosz, “Der Einfluss der Aberrationen aud die optische Ubertragungsfunktion bei kleinen Orts-Frequenzen”, *Opt. Acta* 10, 1 (1963).

[5] M. Born, and E. Wolf, *Principles of Optics*, 7th ed. (Cambridge University Press, Cambridge UK, 1999).

[6] V. N. Mahajan, “Zernike annular polynomials and optical aberrations of systems with annular pupils,”, *Appl. Opt*. 33, 8121 (1994).

[7] B. R. A. Nijboer, *The diffraction theory of aberrations*, Ph.D. thesis, University of Groningen, 1942.

[8] R. J. Noll, “Zernike polynomials and atmospheric turbulence”, *J. Opt. Soc. Am.* 66, 207 (1976).

[9] D. Malacara, *Optical Shop Testing*, 3rd ed. (Wiley, Hoboken, NY, USA, 2007).

[10] S. R. Restaino, S.W. Teare, M. DiVittorio, C. Gilbreath, and D. Mozurkewich, “Analysis of the Naval Observatory Flagstaff Station 1-m telescope using annular Zernike polynomials”, *Opt. Eng.* 42, 2491–2495 (2003).

[11] J. A. Díaz, and R. Navarro,” Orthonormal polynomials for elliptical wavefronts with an arbitrary orientation” , *Appl. Opt.* 53, 2051 (2014).

[12] W. Swantner, and W. W. Chow, “Gram–Schmidt orthogonalization of Zernike polynomials for general aperture shapes,”, *Appl. Opt.* 33, 1832 (1994).

[13] R. Upton and B. Ellerbroek, “Gram-Schmidt orthogonalizationof the Zernike polynomials on apertures of arbitrary shape”, *Opt. Lett.* 29, 2840 (2004).

[14] V. N. Mahajan, “Orthonormal aberration polynomials for anamorphic optical imaging systems with rectangular pupils”, *App. Opt.* 49, 6924 (2010).

[15] V. N. Mahajan, “Zernike annular polynomials for imaging systems with annular pupils”, J. *Opt. Soc. Am.* 71, 75 (1981).

[16] J. A. Díaz, and V. N. Mahajan,” Orthonormal aberration polynomials for optical systems with circular and annular sector pupils”, *Appl. Opt.* 52, 1136 (2013).

[17] F. Liu, B. M. Robinson, P. J. Reardon, and J. M. Geary, “Analyzing optics test data on rectangular apertures using 2-D Chebyshev polynomials”, *Opt. Eng.* 50, 043609 (2011).

[18] V. N. Mahajan, *Optical Imaging and Aberrations, Part III: Wavefront Analysis* (SPIE, 2013).

[ 19 ] G.-m. Dai and V. N. Mahajan, “Nonrecursive orthonormal polynomials with matrix formulation”, *Opt. Lett*. 32, 74 (2007).

[20 ] L. Giraud, and J. Langou, “Robust selective Gram-Schmidt reorthogonalization”, *SIAM J. Sci. Comput*. 25, 417 (2003).

[21] W. Hoffmann, Iterative algorithms for Gram–Schmidt orthogonalization”, *Computing* 41, 353 (1989).

[22] V. N. Mahajan, and G. M. Dai, “Orthonormal polynomials in wavefront analysis: analytical solution,”*J. Opt. Soc. Am. A* 24, 2994 (2007).

[23] B. O'Neill, *Elementary differential geometry*, Rev. 2nd ed. (Academic Press, New York, NY, 2006).

[ 24 ] J. L. Rayces, “Least-squares fitting of orthogonal polynomials to the wave-aberration function”, *Appl. Opt.* 31, 2223 (1992).

[25] D. Malacara, ed., *Optical Shop Testing*, 3rd ed. ( John Wiley & Sons, Hoboken, NJ, 2007).